\documentstyle[preprint,revtex]{aps}
\begin{document}

\draft
\preprint{Submitted to Phys. Rev. Lett. Apr. 15, 1993}
\begin{title}
FERMI-EDGE SINGULARITY WITH BACKSCATTERING IN THE LUTTINGER-LIQUID
MODEL
\end{title}
\author{N. V. Prokof'ev $^{\dag}$}

\begin{instit}
Physics Department, University of British Columbia, 6224 agricultural
RD.,
\end{instit}
\begin{instit}
Vancouver, BC, Canada V6T 1Z1
\end{instit}
\begin{abstract}
We study the response of the interacting electron gas in one dimension
on the local external potential. In the low frequency limit the
power-law singularities are essentially modifyed by
 backscattering effects which, in the case of zero forward
scattering, result in
the universal  critical exponent depending only on the Luttinger-liquid
 interactions. The results obtained may be used to describe singular
responses of the 1D boson chains.

\vspace{1cm}
\noindent
$^{\dag}$ - on leave from: $\;\;\;\;$
Kurchatov Insitute, Kurchatov Square, Moscow 123182, Russia

\end{abstract}


In recent years there has been large interest in non-Fermi-liquid
properties of interacting electron systems in one dimension (1D).
The goal is that the fermion properties  in 1D can be
described precisely even in the presence of electron-electron
interactions \cite{Mattis,Dzya,Hald81}, and one can think about
 possible generalizations of the Luttinger-liquid (LL) state to
higher dimensions \cite{And90}. Also,
recently it has become possible to fabricate really 1D electron
systems in
GaAs inversion layers \cite{Meirav}. The study of the
low frequency x-ray response
provids information both about the spectral density of the electron
 system near the Fermi surface and its reaction on time dependent
 external perturbation.

The interaction between the electrons and a deep core level has strong
influence on the photoemission and soft-x-ray absorption and emission
spectra, $I(\omega )$, \cite{Mahan,And67,Nozieres}. It was found that
 the threshold singularity has the form $I(\omega )\sim A(\omega )\:
\omega^\alpha$, where $A(\omega )$ is the electron spectral density
 near the Fermi surface. The anomalous exponent, $\alpha =
\alpha_{ort}+
\alpha_{ex} $, has two contributions originating from the Fermi-liquid
"orthogonality catastrophe" and
emitted (absorbed) electron rescattering on a core
(the so called orthogonality and excitonic correlations).
In the case of a free electron gas both $\alpha_{ort} $ and
$\alpha_{ex} $
depend on the strength of the core potential.

In the recent papers \cite{Chen92,Ogawa}
the x-ray response of the LL was solved
for the model forward-scattering potential. In the boson representation
\cite{Hald81} the Luttinger-liquid Hamiltonian, $H_L$, and the
forward-scattering interaction have the form $(\hbar =1)$
\begin{equation}
H_L=\sum_{q} \omega_q\;b^+_q\:b_q \;+{1\over 2}{\pi \over L}(v_NN^2+
v_JJ^2)\;,
\label{1}
\end{equation}
\begin{equation}
H_{int}= V e^\varphi \sum_{q} \sqrt{{\vert q\vert \over 2\pi L}}
(b_q+b^+_q)\; ,
\label{2}
\end{equation}
where $\omega_q \approx v_S \vert q\vert $ in the small momentum limit,
 $V$ is
the forward scattering potential, $L \to \infty$ is the length of
 the 1D chain,
$N$ and $J$ are integers describing the
number of right-  and left-moving electrons
with respect to a ground state, $N=\sum_p N_p \;-N_0$,
$J=\sum_ppN_p$ ( right- and left-moving electrons
are labeled by $p=\pm 1$),
and the velocities associated with the
 charge and current excitations
depend on the sound velocity, $v_S$, and the
electron-electron interaction, $v_N=v_S e^{-2\varphi}$,
$v_J=v_S e^{2\varphi}$.
The free electron gas corresponds to $\varphi =0$. In the case of
repulsive interaction between the electrons  the coupling
parameter is negative, $\varphi <0$.
The electron creation operator  can be expressed in the boson basis
as
\cite{Hald81}
\begin{equation}
\Psi_p^+={1\over \sqrt{2\pi \Lambda}}{\bf U}_p
\exp \left\{ \sum_q \sqrt{{2\pi \over L\vert q \vert }}C_{pq}(b_q-b_q^+)
\right\} \;,
\label{3}
\end{equation}
\begin{equation}
C_{pq}=\theta (pq) \cosh \varphi - \theta (-pq) \sinh \varphi \equiv
\left\{
\begin{array}{ll}
\cosh \varphi,\;\;\;\;pq>0 \\
\sinh \varphi,\;\;\;\;pq<0
\end{array} \right.
\label{4}
\end{equation}
where ${\bf U}_p$ is the unitary operator raising $N_p$ and $J_p$ by
one, $\Lambda^{-1}$
is the normalization factor ( or the large momentum cutoff).

When excluded backscattering the authors  of \cite{Chen92,Ogawa}
thus omitted
all possible renormalizations of the scattering amplitude and obtained
an exactly solvable model with the result
\begin{equation}
\alpha_{ort}={1\over 2} g^2 e^{2\varphi} ;
\;\;\;\;\alpha_{ex}= g \;;
\label{5}
\end{equation}
depending explicitely on the dimensionless
potential strength, $g=V/\pi v_S$. However, it was shown
in \cite{Kane} that electrons with repulsive interactions in LL
are completely reflected at zero temperature even by the weakest
impurity potential if the backscattering term is included into
consideration. Using the representation (\ref{3})
this  can be shown by direct calculation of the average $\langle\;
\Psi ^+_{-p}\Psi_p\:\rangle $ which scales as $L^{(1-e^{2\varphi})}$ at
zero temperature. Below the characteristic energy scale, $\omega _B$,
estimated as $(\omega_B /\epsilon_F )^{(1-e^{2\varphi})}\sim g_B$
(where $g_B$ is the bare value of the $2k_F$ scattering amplitude) the
system starts to be decoupled into two semi-infinite lines.
 Thus, in a real system the result (\ref{5}) is not valid
in the low frequency limit where the renormalized backscattering from
the core potential  is dominating. Instead, in this region one can
expect that the exponent $\alpha$ has some universal form and is
defined by the Luttinger-liquid parameters alone. In this paper we
calculate the power-law anomaly in the x-ray spectra in the region
where the core potential results in a perfect reflection of
excitations. It turns out that the universal answer exist only for
the backscattering model. In the general case all scattering processes
contribute to $\alpha$, and the result can not be presented as a sum of
forward and backward exponents.

Our approach is based on the use of canonical transformation relating
eigenfunctions of the system with and without the external perturbation.
Within the golden rule approach the cross section for the x-ray process
with the energy $\omega $ as measured from the thershold is proportional
to
\begin{equation}
I(\omega )\sim 2\pi \sum_{if} \rho^{eq}_i \vert M_{if}\vert ^2 \delta (
E_i+\omega -E_f)\equiv \int^{\infty}_{-\infty} dte^{i\omega t}
\langle\; M^+(t)M(0)\:\rangle \;,
\label{6}
\end{equation}
where $\rho^{eq}_i$ is the equilibrium density matrix of the LL,
$E_i$ and $E_f$ are the energies of the initial and final states,
and $\langle\; \cdots\:\rangle $ stands for the average over the
LL equilibrium. The
electron transfer matrix element has the form
\begin{equation}
M=\Psi^+ e^{S} \;.
\label{7}
\end{equation}
The unitary transformation, $e^{S}$, relating  the eigenstates of
different Hamiltonians $H_L$ and $H_L+H_{int}$ has the advantage of
 calculating all matrix elements in the same basis.

When applyed to the forward scattering interaction, Eq.(\ref{2}), the
canonical
transformation is simply the well known shift of the boson modes
\begin{equation}
e^{S_1}=\exp\left\{ {1 \over 2}ge^\varphi \sum_q
\sqrt{{2\pi \over L\vert q \vert }}(b_q-b_q^+) \right\}\;.
\label{8}
\end{equation}
Now the solution of the x-ray problem is a trivial one because $\Psi^+$
 and
$e^{S_1}$ commute and the exponents are linear functions of the boson
operators. We simply sum the exponents of  $\Psi^+$  and $e^{S_1}$ thus
absorbing the effect of the one-boson final-state interactions into the
new definition of $C_q$
\begin{equation}
C_{q}= \left\{
\begin{array}{ll}
{1\over 2} ge^\varphi + \cosh \varphi,\;\;\;\;q>0 \\
{1\over 2} ge^\varphi - \sinh \varphi,\;\;\;\;q<0
\end{array} \right.
\label{new}
\end{equation}
Formally, we get the problem without the external potential and
\begin{equation}
I(t)=\langle\; M^+(t)M(0)\:\rangle = {1 \over 2\pi \Lambda}
{-i^{\beta_{F}}
(\pi T/\Lambda )^{\beta_{F}} \over \sinh^{\beta_{F}} (\pi Tt)}\;,
\label{9}
\end{equation}
\begin{equation}
\beta_{F}=C^2_{q>0}+C^2_{q<0}=\cosh 2\varphi + 1/2g^2e^{2\varphi}+g\;.
\label{FS}
\end{equation}
 Extracting from this expression the
contribution due to singular spectral density in the LL, $
\alpha = \beta_{F} - \cosh 2\varphi $,
we obtain the result (\ref{5}).

Suppose now that the core potential has only  the backscattering
term which
at sufficiently low energies leads to the perfect reflection of
excitations. In this energy range and $T=0$ we can try to replace
 the original potential  by any perturbation which decouples the LL
state into two semi-infinite chains. In the boson representation  the
perfect reflection of excitations at $q\to 0$ can be obtained by
introducing
a pinning potential at the origin, $U\propto u^2(0)$, where $u(0)$ is
the displacement operator
\begin{equation}
H_{int}={1 \over 2} \sum_{qq^\prime} B_qB_{q^\prime}(b_q+b^+_{q})
(b_{q\prime}+b^+_{q^\prime})\;,
\label{10}
\end{equation}
Here $B_q=(2L\omega_q\epsilon )^{-1/2}$, and $\epsilon^{-1/2} \propto
 v_s$ is
an arbitrary parameter.

Below we consider the zero temperature case which is sufficient for
determining the critical exponent of the Fermi-edge singularity.
 The ground state of the interacting
Hamiltonian can be writtten in the form \cite{Kagan89}
\begin{equation}
\Psi_G=e^{S_2} \Psi_G^{(0)} \equiv {1\over R} \exp \left\{
-{1 \over 2L} \sum_{qq^\prime}
 {A_{qq^\prime} \over \omega_q+\omega_{q^\prime}} b^+_{q} b^+_{q^\prime}
\right\} \Psi_G^{(0)}\;,
\label{11}
\end{equation}
where $R$ is the normalization factor. This expression is valid with
macroscopic accuracy and describes the expansion of the
interacting ground state in terms of the unperturbed wave functions.
The coefficients $A_{qq^\prime}/(\omega_q+\omega_{q^\prime})$ are
an exact amplitudes of finding the state with two bosons $q$ and
$q^\prime$
in $\Psi_G$.  Substituting Eq.(\ref{11}) into the Schrodinger equation
defined by the sum of Hamiltonians (\ref{1}) and (\ref{10})
 we find the following equation
for  $A_{qq^\prime}$
\begin{equation}
 A_{qq^\prime}=B_q B_{q^\prime}-\sum_p  {B_qB_pA_{pq^\prime} \over
\omega_p+\omega_{q^\prime}} - \sum_{p^\prime}
{A_{qp^\prime}B_{p^\prime}B_{q^\prime} \over \omega_q+\omega_{p^\prime}}
 +\sum_{pp^\prime} {A_{qp^\prime}B_{p^\prime}B_{p}
A_{pq^\prime} \over (\omega_q+\omega_{p^\prime}) (\omega_p+
\omega_{q^\prime})}
\;.
\label{12}
\end{equation}
The corresponding terms in Eq.(\ref{12}) tell us that the state with
 two bosons
can be obtained either by direct creation of one boson pair
in $\Psi_G^{(0)}$,
or by rescattering of the existing pair, or by absorption of one
pair in the two pair state. Obviously, the solution has a separable
 form
\begin{equation}
A_{qq^\prime}=(B_q\nu_q)(B_{q^\prime}\nu_{q^\prime})\;;\;\;\;
\nu_q=1-\nu_q \sum_p {B_p^2 \nu_p \over \omega_p+\omega_{q}}\;.
\label{13}
\end{equation}
We can transform this equation into the linear one substituting $\nu_p$
in the
r.h.s. by Eq.(\ref{13}) itself and regrouping the terms. The result is
\begin{equation}
\nu_q=1-2\nu_q \sum_p {B_p^2 \omega_p \over \omega_p^2-\omega_{q}^2}+
 \sum_p {B_p^2 \nu_p \over \omega_p-\omega_{q}}\;.
\label{14}
\end{equation}
This equation is well known in the theory of singular integral equations.
 However, we can easily find the behavior of $\nu_q$ at
small $q$ simply noting that the sum in Eq.(\ref{13}) is divergent  at
 small
$\omega_p$ and $\nu_{q\to 0} \to 0$. With this property of the solution
we substitute the low-energy anzats $\nu_q = (c\omega_q)^{1/2}$ into
Eq.(\ref{13}) and find
\begin{equation}
c=\left( {1 \over 2\pi L\epsilon } \sum_p {1 \over \omega_p^{1/2} (
\omega_p + \omega_q)} \right)^{-1/2}=\sqrt{2\epsilon v_S}\;.
\label{15}
\end{equation}
Thus, in the low-frequency range the amplitudes acquire constant values
$A_{qq^\prime} \approx v_S$. Then, we can write the canonical
transformation in the  form
\begin{equation}
e^{S_2} = {1\over R} \exp \left\{
-{1 \over 2L} \sum_{qq^\prime}
 {v_S \over \omega_q+\omega_{q^\prime}} b^+_{q} b^+_{q^\prime} \right\}\;.
\label{16}
\end{equation}
The integrals determining the x-ray response are infrared divergent,
that is why we can restrict ourselves by the low-frequency solution.

To calculate the normalization factor from $\langle \: (e^{S_2})^+
e^{S_2} \: \rangle  =1$,
 we make use of the
well known trick substituting $S_2$ by $\lambda S_2$ and then
differentiating the above formula over $\lambda $
\begin{equation}
\ln\: R^2 = -\int_{0}^{1} d\lambda {v_S \over 2L}  \sum_{qq^\prime}
 {1 \over \omega_q+\omega_{q^\prime}} \langle \: (e^{S_2})^+\;
 e^{\lambda S_2 }
\;b_q^+b_{q^\prime}^+ \: \rangle _{_{_{\bf C}}} \;.
\label{17}
\end{equation}
Now we have to calculate only the sum of connected diagrams
in Eq.(\ref{17}). The $n$-th order diagram is generated by the $n$-th
 term
in the series expansion of $e^{\lambda S_2}$ with the result
\begin{equation}
\ln\: R^2 = {1 \over 2}\sum_{n=1}^{\infty}{1 \over n}
\sum_{q_1 \cdots q_{2n}}{(v_S/L)^{2n} \over (\omega_1+\omega_2)
(\omega_2+
\omega_3)\cdots (\omega_{2n-1}+\omega_{2n})(\omega_{2n}+\omega_1)}\;.
\label{18}
\end{equation}
Extracting the logarithmic singularity we can present the result in
 the form
\begin{equation}
\ln\: R^2 = {1 \over 8\pi^2}\sum_{n=1}^{\infty}{4^n (n-1)! (n-1)! \over
n(2n-1)! }\; \int_{q_{min}}^{q_{max}} {dq \over q} \equiv {1 \over 8}
\ln \left( {Lq_{max}\over 2\pi} \right) \;.
\label{19}
\end{equation}
The dynamic response $D(t)= \langle \:
 (e^{S_2(t)})^+ e^{S_2} \: \rangle$, or the core Green function, is
 defined
by the same sum of diagrams as  in Eq.(\ref{18}) with additional factors,
$\exp \{ -i(\omega_1+\cdots +\omega_{2n})t\} -1$, which
effectively cut off the low-energy divergency at $1/t$. Thus, in the
long-time asymptotics we have
\begin{equation}
D(t)\sim t^{-1/8}\;.
\label{20}
\end{equation}

We find that the reaction of the 1D boson chain on the local
pinning potential with perfect reflection  of low-energy  excitations
is infrared divergent. Note, that the same result comes if the
interaction, Eq.(\ref{10}), is replaced by the perturbation induced by an
infinite boson mass ($M\to \infty $) at the origin
\begin{equation}
H_{int}=\left( {1\over M } - {1\over m} \right) {P^2(0) \over 2} \equiv
-{P^2(0) \over 2m} \;,
\label{21}
\end{equation}
where $P(0)$ is the boson momentum operator. The  canonical
transformation in this case has exactly the same form, Eq.(\ref{16}), in
the $q\to 0$ limit.

Compare now the result obtained with the well known expression for the
normalization exponent in a free electron gas \cite{And67}
\begin{equation}
\alpha_{ort}=\sum_{l=0,1} \left( {\delta_l \over \pi } \right) ^2
 ={1\over 8}\;,
\label{22}
\end{equation}
where $l=1,0$ is the index of even and odd phase shifts.
In the case of perfect reflection the constraint, $\delta_0-\delta_1=
\pi /2+\pi n$, is imposed on the phase shifts. Then, from Eq.(\ref{22})
 we conclude
 that backscattering is described by two nonzero values $
\delta_0=-\delta_1=\pm \pi /4$. Indeed, the backscattering potential
 when
expressed in terms of even and odd parity combinations,
$V_{l=0,1}={1\over 2}(V(0)\pm V(2k_F))=\pm {1\over 2}V(2k_F)$,
 generates
two phase shifts $\delta_0=-\delta_1$. This case must be
contrasted to
the electron scattering by strong $\delta$-functional potential
when $\delta_0=\pm \pi /2$, $\delta_1=0$.

Calculate now the x-ray response, Eq.(\ref{6}), with the transfer matrix
given by Eqs.(\ref{3}) and (\ref{16}). We note that all $\underline{
connected}$ diagrams involving one-boson processes from $\Psi^+ $ and
 $\Psi $ contain only two one-boson vertecies as the starting and
ending lines. Thus, all possible connected diagrams are given by
(compare with Eq.(\ref{18}))
\begin{eqnarray}
\ln\: I(t)&=&\ln\: D(t) + D_{el}(t)+D_1(t)\;,\\
D_{el}(t)&=&(C^2_{q>0}+C^2_{q<0})
 \: \int {dq \over q} (e^{-i\omega_qt}-1) \sim -\cosh\:2\varphi
\:\ln\:t\;,\\
D_1(t)&=&-{\pi \over 2}(C_{q>0}+C_{q<0})^2
\sum_{n=2}^{\infty} (-1)^n\sum_{
q_1\cdots q_n}{(v_S/L)^n\phi_1(t)\phi_2(t)e^{-i(\omega_3+\cdots
+\omega_n)t}
\over \sqrt{\omega_1\omega_2}(\omega_2+\omega_3)\cdots (\omega_n
+\omega_1)}\;.
\label{25}
\end{eqnarray}
Here in the last expression $\phi_q=e^{-i\omega_qt}-1$.
As expected, the second term describes the single particle Green
function
for the conduction band and is an intrinsic property of  LL. The
only contribution in the sum (25) which is infrared divergent as $t\to
\infty$
is that for $n=2$. Summing up all logarithmic contributions we find
the universal exponent describing the x-ray singularity in the case of
backscattering potential
\begin{equation}
\beta_B =C^2_{q>0}+C^2_{q<0} +{1\over 8} + {1\over 2} (C_{q>0}+C_{q<0})^2
\equiv \cosh 2\varphi +
{1\over 8} +{1\over 2}e^{-2\varphi} \;.
\label{26}
\end{equation}

In the general case both the forward and backscattering processes
contribute
to the x-ray response. Let the interaction Hamiltonian to be the sum
of Eq.(\ref{2}) and Eq.(\ref{10}). The ground state of the system  can be
obtained by
switching  on adiabatically the two-boson scattering first,
 Eqs.(\ref{11}) and
(\ref{16}), and introducing the one-boson interaction afterwards.
 Due to the
fact that the Hamiltonians (\ref{10}) and (\ref{21}) are quadratic
in the boson
field, the one-boson transformation $e^{S_1}$, Eq.(\ref{8}), is not
affected
by the first unitary rotation of boson variables. Thus,
\begin{equation}
e^{S}=e^{S_1}\:e^{S_2} \;.
\label{27}
\end{equation}
As before, the one-boson exponent can be combined with the
$\Psi^+ $ operator (see Eq.(\ref{new})), and formally we get the same
 kind of
 problem as in the case of backscattering potential alone.
The final result has the form
\begin{equation}
\beta =\cosh 2\varphi + (g^2e^{2\varphi}+2g)+(
{1\over 8} +{1\over 2}e^{-2\varphi}) \;.
\label{28}
\end{equation}
At first sight the expression is the sum of two independent
contributions from the forward and backscattering processes. However,
this is not the case. The forward scattering term in Eq.(\ref{28}) is
\underline{twice} the
result (\ref{5}) obtained when neglecting the $2k_F$ momentum transfer
from the potential.

Up to this point we considered the spinless case. Due to the spin-charge
separation in 1D, we can include spin excitations into the consideration
by introducing another boson field, $d_q$. For the spin-independent
scattering potential this additional boson field has no forward
scattering from the core level. Thus the result (\ref{28}) is
modifyed to $\beta = \beta_{spin}+\beta_{charge}$
\begin{eqnarray}
\beta_{spin}&=&{1\over 2} + ({1\over 8}+{1\over 4})\;;  \nonumber \\
\beta_{charge}&=&{1\over 2}\cosh 2\varphi
+(g^2e^{2\varphi}+\sqrt{2} g) + ({1\over 8}+{1\over 4}e^{-2\varphi})\;
\end{eqnarray}
($\beta = \cosh ^2\varphi + {1\over 2} (g^2e^{2\varphi}+\sqrt{2}g) $
 without backscattering \cite{Ogawa}).
In the framework of the LL model the forward scattering phase shift
is proportional to the potential strength $\delta_{FS} = -\pi g$
\cite{Schotte} due to the infinite energy spectrum with the
 linear dispersion relation. Assuming that the LL model describes only
the low-energy physics of the real electron system we have to
consider the parameter $\pi g$ as the phase shift rather than a bare
potential. This is important in understanding that $g$ in Eq.(\ref{28})
 can
not be arbitrary large.

We assumed in our calculation that the backscattering processes resulted
in a perfect reflection of excitations from the potential in the
 low-energy
region as confirmed by scaling arguments and the exact solution of the
special case $e^\varphi = 1/2$ \cite{Kane}. This means that our results
are valid only below the characteristic frequency $\omega_B$. It is not
clear at the moment to what extent the model description of the
perfect reflection at $\omega \to 0$ by pinning potential (or
by infinite mass) at the origin is adequate to the problem. One can
argue that the two-boson interaction is the simplest one of the kind,
 and
the result obtained is independent of the strength of the potential.
Two nonzero phase shifts equal to $\vert \pi /2 \vert $ are in
agreement with the result expected for the electron reflection in the
absense
of forward scattering.

Obviously, these results can be applyed for studing the singular behavior
of the one-dimensional boson systems.


\begin{references}
\bibitem{Mattis} Mattis D.C., and E.H. Lieb, J. Math. Phys., {\bf 6},
304 (1965).
\bibitem{Dzya} Dzyaloshinskii I.E , and Larkin A.I. , Zh. Eksp. Teor.
Fiz.
{\bf 65}, 411 (1974) [Sov. Phys. JETP {\bf 38}, 202].
\bibitem{Hald81} Haldane F.D.M., J. Phys. {\bf C}, {\bf 14}, 2585 (1981).
\bibitem{And90} Anderson P.W.,
Phys.\ Rev.\ Lett.,\ {\bf 64}, 1839 (1990).
\bibitem{Meirav} Meirav U., M.A. Kastner, M. Heiblum, and S.J. Wind,
 Phys. Rev. B {\bf 40}, 5871 (1989).
\bibitem{Mahan} Mahan G.D., {\it Many-Particle Physics}, (Plenum,
 New York, 1981).
\bibitem{And67} Anderson P.W.,
Phys.\ Rev.\ Lett.,\ {\bf 18}, 1049 (1967).
\bibitem{Nozieres} Nozieres P.,  and C.T. DeDominicis, Phys. Rev.,
 {\bf 178}, 1097 (1969).
\bibitem{Chen92} Lee D.K.K.,  and Y. Chen,
Phys.\ Rev.\ Lett.,\ {\bf 69}, 1399 (1992).
\bibitem{Ogawa} Ogawa T., A. Furusaki, and N. Nagaosa, Phys. Rev. Lett.,
{bf 68}, 3638 (1992).
\bibitem{Kane} Kane C.L., and M.P.A. Fisher, Phys.\ Rev.\ Lett.,\
{\bf 68}, 1220 (1992); Preprint (1992).
\bibitem{Kagan89}  Kagan Yu. and N.V. Prokof'ev, Zh. Eksp. Theor.
68 Fiz. {\bf 96}, 2209 (1989)  [Sov. Phys. JETP {\bf 69}, 1250].
\bibitem{Schotte} Schotte K.D., and U. Schotte, Phys. Rev. {\bf 182},
479 (1969).
\end{references}
\end{document}